\documentclass[twocolumn,showpacs,aps,prb,superscriptaddress]{revtex4-1}
\usepackage{graphics}
\usepackage{color}
\newcommand{\textdegree}{\ensuremath{^\circ}}
\begin{document}


\title{Magnetic structures of Mn$_{3-x}$Fe$_{x}$Sn$_{2}$: an experimental
and theoretical study}

\author{Q. Recour}
\affiliation{Institut Jean Lamour, D\'epartement P2M, CNRS (UMR 7198), Universit\'e
de Lorraine, BP 70239, 54506 Vand\oe uvre-l\`es-Nancy Cedex, France}
\author{ V. Ban}
\affiliation{Institut Jean Lamour, D\'epartement P2M, CNRS (UMR 7198), Universit\'e
de Lorraine, BP 70239, 54506 Vand\oe uvre-l\`es-Nancy Cedex, France}
\affiliation{Institue of Condensed Matter and Nanosciences, Universit\'e Catholique de Louvain, Place L. Pasteur 1, 
B-1348 Louvain-la-Neuve, Belgium}
\author{ Z. Gercsi}
\affiliation{Department of Physics, Blacklett Laboratory, Imperial College London,
London SW7 2AZ, United Kingdom}
\author{T. Mazet}
\email[correspondence:]{thomas.mazet@univ-lorraine.fr}
\author{ M. Fran\c cois} 
\author{ B. Malaman}
\affiliation{Institut Jean Lamour, D\'epartement P2M, CNRS (UMR 7198), Universit\'e
de Lorraine, BP 70239, 54506 Vand\oe uvre-l\`es-Nancy Cedex, France}

\date{\today}

\begin{abstract}
We investigate the magnetic structure of Mn$_{3-x}$Fe$_{x}$Sn$_{2}$
using neutron powder diffraction experiments and electronic structure
calculations. These alloys crystallize in the orthorhombic Ni$_{3}$Sn$_{2}$
type of structure ($Pnma$) and comprise two inequivalent sites for
the transition metal atoms (4$c$ and 8$d$) and two Sn sites (4$c$
and 4$c$). The neutron data show that the substituting Fe atoms predominantly
occupy the 4$c$ transition metal site and carry a lower magnetic
moment than Mn atoms. Four kinds of magnetic structures are encountered
as a function of temperature and composition: two simple ferromagnetic
structures (with the magnetic moments pointing along the $b$ or $c$
axis) and two canted ferromagnetic arrangements (with the ferromagnetic
component pointing along the $b$ or $c$ axis). Electronic structure
calculations results agree well with the low-temperature experimental
magnetic moments and canting angles throughout the series. Comparisons
between collinear and non-collinear computations show that the canted
state is stabilized by a band mechanism through the opening of a hybridization
gap. Synchrotron powder diffraction experiments on Mn$_{3}$Sn$_{2}$
reveal a weak monoclinic distortion at low temperature ($\beta$$\sim$90.08\textdegree{} at 175 K). 
This lowering of symmetry could explain the
stabilization of the $c$-axis canted ferromagnetic structure, which
mixes two orthorhombic magnetic space groups, a circumstance that
would otherwise require unusually large high-order terms in the 
spin Hamiltonian. 
\end{abstract}

\maketitle

\section{Introduction\label{sec:Introduction}}

Mn$_{3}$Sn$_{2}$ and its Fe derivatives Mn$_{3-x}$Fe$_{x}$Sn$_{2}$
($x$$<$1) exhibit an uncommon two-peak magnetocaloric response in the
200-300 K temperature range\cite{Mn3Sn2,Fe,Cp} which could be useful in future near
room temperature magnetic cooling applications.\cite{Rev1,Rev2}
These orthorhombic phases are isotypic with Ni$_{3}$Sn$_{2}$ ($Pnma$)
and comprise two inequivalent crystallographic sites for both the
transition metal (4$c$ and 8$d$) and the tin atoms (4$c$ and 4$c$).
\cite{cryst,PRB} 

It has been shown that Mn$_{3}$Sn$_{2}$ undergoes
three distinct magnetic transitions. \cite{Mn3Sn2,PRB} At $T_{C1}$$\sim$262
K, the Mn 4$c$ site orders ferromagnetically, with the Mn moments
pointing along the $b$ axis, and polarizes the Mn 8$d$ magnetic
moments. The Mn 8$d$ sublattice cooperatively orders at $T_{C2}$$\sim$227
K in a canted ferromagnetic arrangement with the ferromagnetic component
along the $b$ axis and the antiferromagnetic one along the $a$ axis,
while the ferromagnetic order of Mn 4$c$ is unaltered. These two
ferromagnetic-like second order transitions yield the two-peak magnetocaloric
response. Finally, below $T_{t}$$\sim$197 K the ferromagnetic
component of both Mn 4$c$ and Mn 8$d$ abruptly reorients from the
$b$ axis to the $c$ axis, the canting angle ($\alpha$$\sim$51\textdegree)
of Mn 8$d$ being unchanged
at the transition. Based on group
theory arguments, it has been concluded that the magnetic structure
below $T_{t}$ either requires unusually large high-order terms in the spin
Hamiltonian or a (undetected) monoclinic distortion.\cite{PRB} The low-temperature
magnetic moments carried by Mn atoms are rather large, close to $\sim$2.3
and $\sim$3.0 $\mu_{B}$ for Mn 4$c$ and Mn 8$d$, respectively.
From magnetization data,\cite{Fe} it has been observed that the Fe
for Mn substitution in Mn$_{3-x}$Fe$_{x}$Sn$_{2}$, which is limited
to about $x$$\sim$0.9, alters the magnetic behavior but leaves the
magnetization roughly constant around 5.4 $\mu_{B}$/f.u. throughout
the series. The alloys with low Fe content ($x$$\leq$0.4) have transition
temperatures very close to that of Mn$_{3}$Sn$_{2}$. A further increase
in the Fe content yields an increase of $T_{C1}$ up to $\sim$290
K in $x$=0.9 as well as a decrease of $T_{C2}$ and $T_{t}$
down to $\sim$171 K and $\sim$164 K, respectively. Interestingly,
for the intermediate compositions ($x$=0.5 and 0.6) only the two
ferromagnetic-like transitions are present. There are further modifications
in the shape of the thermomagnetic curves throughout the series which
suggest that the Fe substitution does not simply result in a shift
of the magnetic transition temperatures.

In this paper, we investigate the magnetic properties of Mn$_{3-x}$Fe$_{x}$Sn$_{2}$
alloys at a microscopic level using neutron powder diffraction experiments
and band structure calculations. Further, we check from synchrotron
diffraction measurements if a weak lowering of symmetry occurs in
Mn$_{3}$Sn$_{2}$.
The organization of the paper is as follows. Section
\ref{sec:Experimental} gives the experimental and computational details.
In Sec. \ref{sec:Neutron}, we present and analyze the neutron powder
diffraction experiments on Mn$_{3-x}$Fe$_{x}$Sn$_{2}$ (0$<$$x$$\leq$0.8).
Sec. \ref{sec:Electronic} gives the results of our collinear and
non-collinear electronic structure calculations on Mn$_{3}$Sn$_{2}$
and Mn$_{3-x}$Fe$_{x}$Sn$_{2}$ as compared with experiments. Sec.
\ref{sec:Synchrotron} deals with temperature dependent X-ray synchrotron
diffraction on Mn$_{3}$Sn$_{2}$. Finally, the paper is summarized
in Sec. \ref{sec:Summary}.

\section{Experimental and computational details\label{sec:Experimental}}

Powder neutron diffraction experiments were carried out at the Institut
Laue Langevin (ILL), Grenoble (France) using the two axis D1B diffractometer
($\lambda$=2.52 $\textrm{\AA}$, 128\textdegree{} position
sensitive detector, step of 0.1\textdegree). The samples
investigated here (with $x$=0.1, 0.2, 0.3, 0.4, 0.5, 0.6, 0.7, and
0.8) were those used in reference~\onlinecite{Fe}. Note that the sample
with $x$=0.9 of ref.~\onlinecite{Fe} was excluded from the present study
because of its high content in the binary FeSn impurity that might
obscure the analysis of the neutron data. Numerous diffraction patterns
were recorded in the 2-300 K temperature range for each alloy using
a standard helium cryostat. The analysis of the neutron data was performed
by Rietveld refinements using the Fullprof software.\cite{FullP}
Due to an imperfect monochromatization of the incident beam, the refinements
were carried out by considering a second harmonic ($\lambda/2$) contamination
of 0.6\% intensity. Besides the peak profile parameters, the refinements
comprised scale factor, zero-shift, cell parameters, crystallographic
positions, overall Debye-Waller factor, magnitude and orientation
of the Mn/Fe moments. The Mn/Fe site occupancy ratio was determined
from room temperature data ($i.e.$ in the paramagnetic state). In order
to reduce the number of intensity dependent parameters, the occupancy
ratios were kept fixed for the fits in the magnetically ordered state.
The nuclear contribution from the impurities (MnSn$_{2}$ and MnO)
was taken into account during the refinements.

Powder synchrotron diffraction experiments were
carried out at the Swiss Light Source (Paul Scherrer Institute, Viligen,
Switzerland) using the MS-powder beam line\cite{MS-line} to check
the occurrence of a possible weak lowering of symmetry. Due to time
limitation, only the Mn$_{3}$Sn$_{2}$ parent compound has been investigated.
Diffraction patterns  were measured in glass
capillaries ($\phi$=0.3 mm) with standard Debye-Scherrer geometry
using a multistrip detector\cite{strip} ($\lambda$=0.7999 $\textrm{\AA}$,
calibrated using silicon standard from NIST). The one-dimensional detector
allows for the measurements of entire diffraction patterns over 120\textdegree{}
in a few seconds. Four diffraction patterns corresponding to four
different positions of the detector were recorded in order to improve
the Bragg intensity/background ratio. The patterns were recorded
from 300 K down to 175 K using a cryo-jet device. The patterns were
refined using the profile matching procedure of the Fullprof software
\cite{FullP} (Lebail decomposition) for extracting the temperature
dependence of the lattice parameters. The lines profile was modeled
with function 7 of Fullprof (Thompson-Cox-Hastings pseudo-Voigt convoluted
with axial divergence asymmetry function). The instrumental function
was determined using a small line-width sample (NAC; Na$_{2}$Ca$_{3}$Al$_{2}$F$_{4}$)
as reference (half-width=0.024\textdegree{} at 2$\theta$=23\textdegree).

The electronic structure calculations were performed using the Vienna
ab initio simulation package (VASP) code, based on DFT within projector
augmented wave (PAW) method\cite{VASP} with Perdew-Burke-Ernzerhof
(PBE) parametrization.\cite{PBE} Site-based magnetic moments were
calculated using the Vosko-Wilk-Nusair interpolation\cite{Vosko}
within the general gradient approximation (GGA) for the exchange-correlation
potential. A $k$-point grid of 13$\times$9$\times$15 was used
to discretize the first Brillouin zone and the energy convergence
criterion was set to $10^{-6}$ eV during the energy minimization
process. The density of states (DOS) plots presented in this work
were calculated on a dense (15$\times$19$\times$17) $k$-grid
for high accuracy. The spin-orbit interaction was turned on during
the non-collinear calculations. 

The minimal, 20 atoms basis cell consists of twelve Mn atoms (situated
on the Mn 4$c$ and Mn 8$d$ sites) and eight Sn atoms (situated on
two distinct 4$c$ positions)  and was used to evaluate the total energies
and magnetic properties of the alloys. The effect of partial replacement
of Mn for Fe on the electronic structure was only considered into
the 4$c$ crystallographic site supported by experimental findings (discussed
in Sec. \ref{sub:Para}). Using this model, the effect of doping was
simulated by the replacement of a manganese atom by iron which represents
an $x$=1/4 compositional change in the Mn$_{3-x}$Fe$_{x}$Sn$_{x}$
formula. The experimental lattice parameters measured at 2 K were
used for the calculations. For the non-collinear spin arrangements,
we performed fully self-consistent calculations with the direction
of the magnetic moments constrained into pre-set alignments (see Sec.
\ref{sec:Electronic} for details). It must be emphasized that the
amplitude of the moments could vary freely during the minimization
process. The symmetry analysis was turned off for the calculations
and the full set of $k$-grid points was used. Finally, the band structure
plots presented in this work were obtained from non-constrained solutions.
This later was achieved from the constrained solutions by stepwise
removal of the constrain from one run to another in order to keep
the solution stable.

\section{Neutron diffraction study\label{sec:Neutron}}

\subsection{Paramagnetic state\label{sub:Para}}

\begin{figure}[t]
\begin{center}
\scalebox{0.75}{\includegraphics{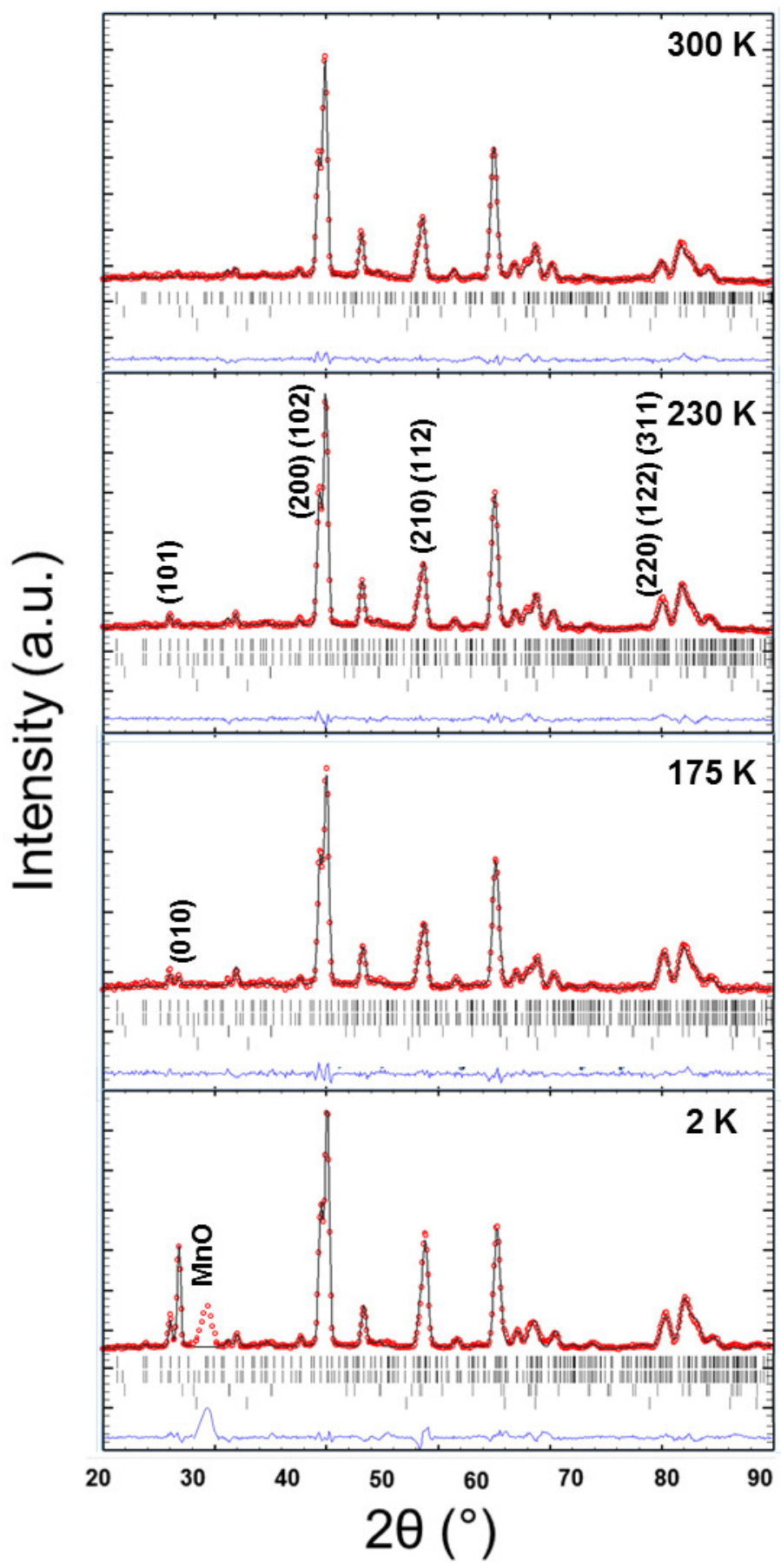}}
\end{center}
\vspace{-0.5cm}
\caption{\label{fig:Neutron}Neutron diffraction patterns of Mn$_{2.2}$Fe$_{0.8}$Sn$_{2}$.}
\vspace{-0.5cm}
\end{figure}

\begin{figure}[h]
\begin{center}
\scalebox{0.35}{\includegraphics{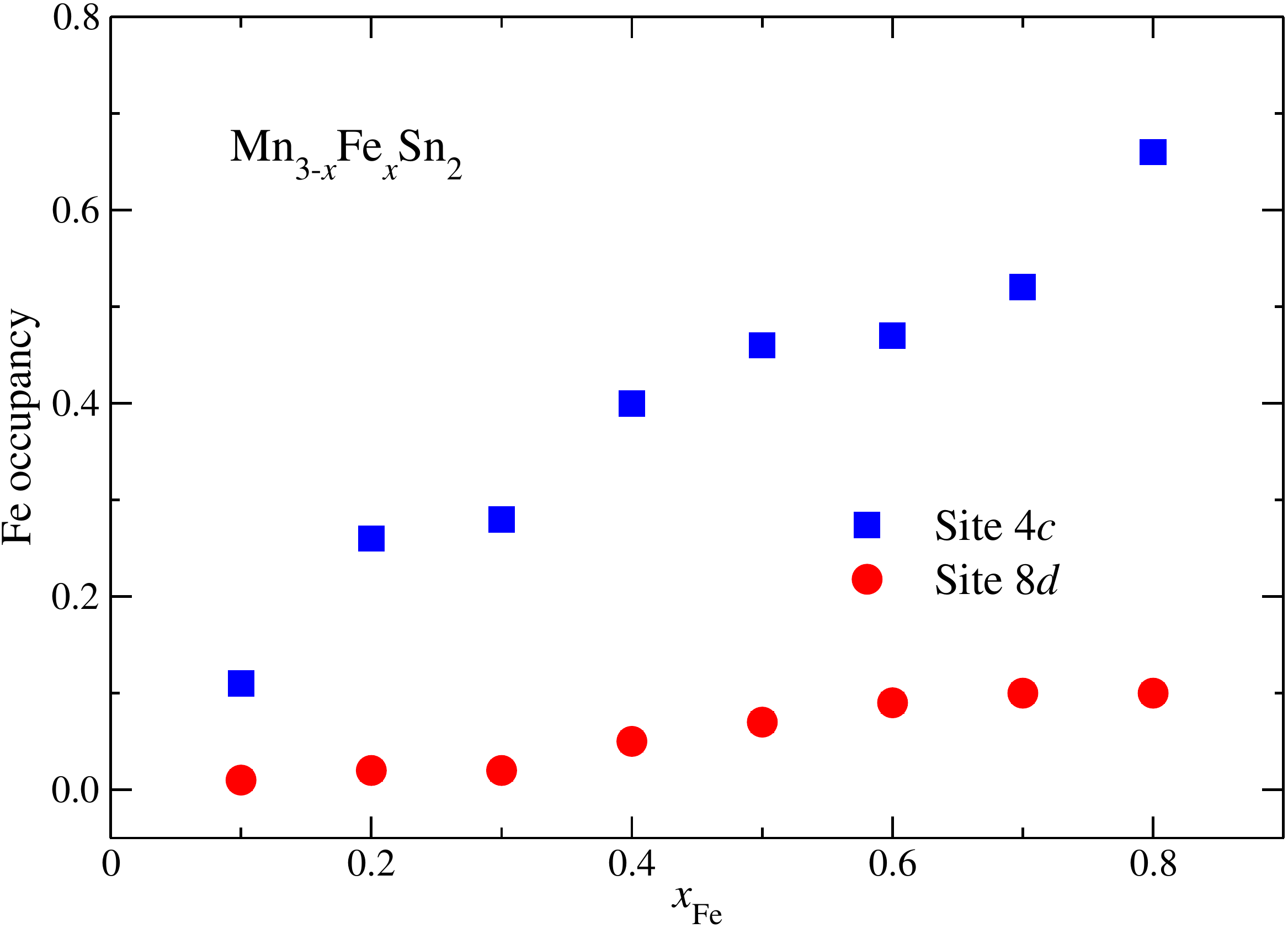}}
\end{center}
\vspace{-0.2cm}
\caption{\label{fig:Occ_Fe}Composition dependence of the occupancy of the
4$c$ and 8$d$ transition metal sites by Fe atoms from neutron data.}
\vspace{-0.5cm}
\end{figure}

In all cases, the neutron diffraction pattern recorded at 300 K ($i.e.$
in the paramagnetic state, see top panel of Figure \ref{fig:Neutron})
comprises nuclear reflections from the Ni$_{3}$Sn$_{2}$-type main
phase as well as tiny diffraction peaks originating from the MnSn$_{2}$
and MnO impurities whose total amount represents a few wt.\%. Refinements
of the nuclear structure yield $R_{N}$ agreement factors close
to 5\%, indicating that no significant deviation from the Ni$_{3}$Sn$_{2}$
type of structure occurs throughout the series. Due to the marked
difference between the bound coherent scattering lengths of Mn and
Fe ($b_{Mn}$=-3.73 fm, $b_{Fe}$=9.45 fm), the Rietveld analysis
of the neutron data allowed us to determine the atomic distribution
of Mn and Fe on both sites available (4$c$ and 8$d$). The evolution
of the Fe atomic distribution as a function of the Fe content is presented
in Figure \ref{fig:Occ_Fe}. It is found that Fe atoms preferentially
occupy the 4$c$ site, although the 8$d$ site begins to be substituted
in the richest Fe compositions.

\subsection{Magnetically ordered state}

\begin{table}[b]
\caption{\label{tab:The-magnetic-modes}The magnetic modes associated with with \textbf{ k}=(0  0  0) for the Wyckoff positions 8$d$ and 4$c$ of the two relevant
orthorhombic magnetic space groups $Pn'm'a$ and $Pn'ma'$ as well as for the Wyckoff position 4$e$ of the monoclinic magnetic space group $P2'_{1}/n'11$ (from references~\onlinecite{Schob,Rousse}). $P2'_{1}/n'11$ is the intersection of  $Pn'm'a$ and $Pn'ma'$. The 8$d$ position of $Pnma$ is split into two 4$e$ positions of $P2_{1}/n11$ and the  4$c$ position of $Pnma$ transforms into 4$e$ position  of $P2_{1}/n11$. The modes are A(+ - - +), C(+ + - -), F(+ + + +), and G(+ - + -).}
\begin{ruledtabular}
\begin{tabular}{cccc}
 Position &  $Pn'm'a$ &$Pn'ma'$&$P2'_{1}/n'11$\\
\colrule
8$d$      & $G_{Bx}$$C_{By}$$F_{Bz}$&$A_{Bx}$$F_{By}$$C_{Bz}$ \\
4$c$      &  $C_{x}$$F_{z}$ &$ F_{y}$         &  \\
4$e$      &               &                 &    $A_{x}$$F_{y}$$F_{z}$ \\
\end{tabular}
\end{ruledtabular}
\end{table}

\begin{figure*}[t]
\begin{center}
\scalebox{0.80}{\includegraphics{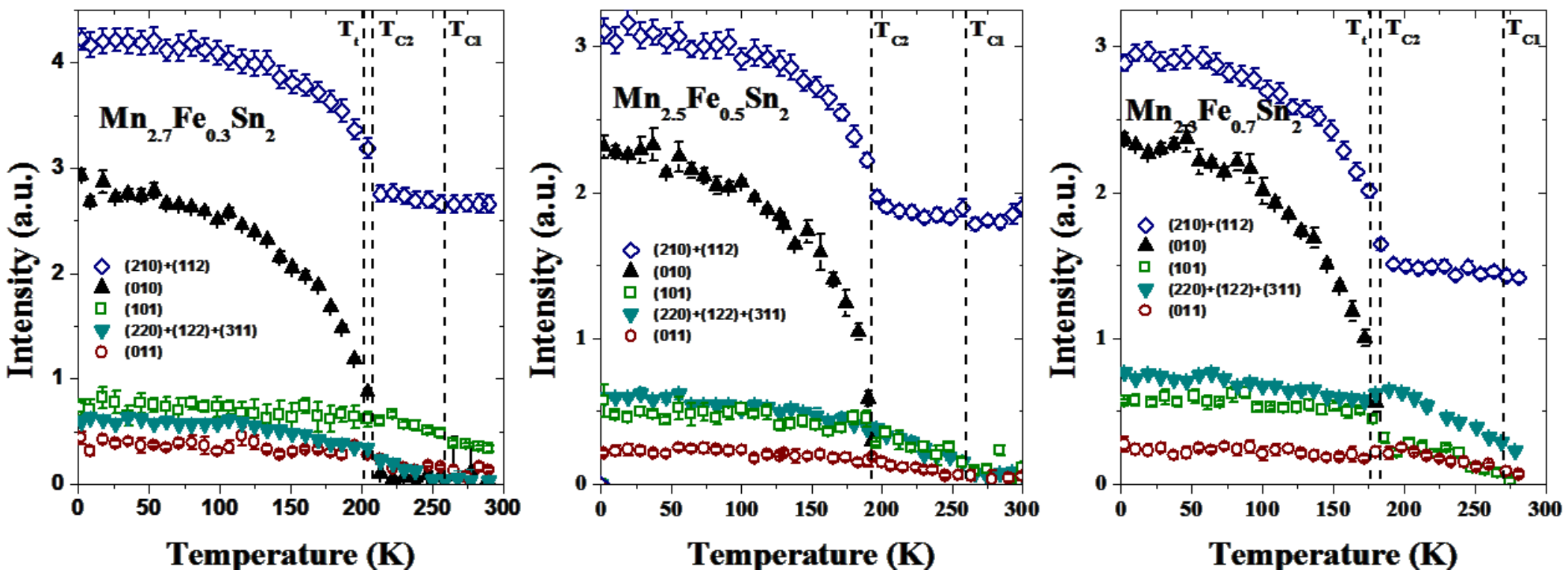}}
\end{center}
\vspace{-0.5cm}
\caption{\label{fig:I=00003Df(T)}Temperature dependence of the intensity of
some selected peaks or groups of peaks for three representative compositions
($x$=0.3, 0.5, and 0.7).}
\vspace{-0.3cm}
\end{figure*}

Refinements of the nuclear structure show that, within the accuracy
of the measurements, the Ni$_{3}$Sn$_{2}$ type of structure ($Pnma$)
is stable down to 2 K. Besides the peaks corresponding to the $Pnma$
space group, the neutron thermodiffractograms 
are characterized by the appearance of the anti-$n$ (010) reflection
below $T_{C2}$ (Figure \ref{fig:Neutron}) and by the growing of a peak at 2$\theta$$\sim$28.7\textdegree{}
due to the antiferromagnetic ordering of the MnO impurity below $T$$\sim$120
K. Hence, in the whole ordered temperature range, the magnetic structures
of Mn$_{3-x}$Fe$_{x}$Sn$_{2}$ alloys are commensurate associated
with the propagation vector \textbf{k}=(0 0 0), similarly to the
binary Mn$_{3}$Sn$_{2}$ parent compound.\cite{PRB} However, there
are important differences in the thermal dependence of the peak intensities
which allow the classification of Mn$_{3-x}$Fe$_{x}$Sn$_{2}$ alloys
into three groups (figure \ref{fig:I=00003Df(T)}): alloys with $x$$\leq$0.4,
alloys with $x$=0.5 and 0.6, and alloys with $x$$\geq$0.7. The magnetic
modes\cite{Bertaut} of the 4$c$ and 8$d$ sites for the two magnetic
space groups of $Pnma$ with\textbf{ k}=(0 0 0) needed for the present
neutron analysis are given in Table \ref{tab:The-magnetic-modes}.

\begin{table*}[t]
\caption{\label{tab:Mom 01 03}Refined magnetic moments for Mn$_{2.9}$Fe$_{0.1}$Sn$_{2}$
and Mn$_{2.7}$Fe$_{0.3}$Sn$_{2}$ from neutron data.}
\begin{ruledtabular}
\begin{tabular}{lcccccccc}
 &  &  & $x$=0.1 &  &  & & $x$=0.3 & \\
 &  & 230 K & 200 K & 2 K &  & 230 K & 205 K & 2 K\\
\colrule
Site 8$d$ & Mode & $F_{By}$ & $A_{Bx}F_{By}$ & $A_{Bx}F_{Bz}$ &  & $F_{By}$ & $A_{Bx}F_{By}$ & $A_{Bx}F_{Bz}$\\
 & $m_{x}(\mu_{B})$ &  & 1.60(1) & 2.15(2) &  &  & 0.68(2) & 2.17(2)\\
 & $m_{y}(\mu_{B})$ & 1.11(4) & 1.36(3) &  &  & 0.87(4) & 1.18(4) & \\
 & $m_{z}(\mu_{B})$ &  &  & 1.78(2) &  &  &  & 1.95(3)\\
\vspace{3mm}
 & $m(\mu_{B}$) & 1.11(4) & 2.10(2) & 2.79(2) &  & 0.87(4) & 1.39(3) & 2.92(2)\\ 
Site 4$c$ & Mode & $F_{y}$ & $F_{y}$ & $F_{z}$ &  & $F_{y}$ & $F_{y}$ & $F_{z}$\\
 & $m_{x}(\mu_{B})$ &  &  &  &  &  &  & \\
 & $m_{y}(\mu_{B})$ & 1.58(5) & 1.92(3) &  &  & 1.44(5) & 1.79(4) & \\
 & $m_{z}(\mu_{B})$ &  &  & 2.22(5) &  &  &  & 2.00(4)\\
\vspace{3mm}
 & $m(\mu_{B}$) & 1.58(5) & 1.92(3) & 2.22(5) &  & 1.44(5) & 1.79(4) & 2.00(4)\\
$R_{mag}$, $R_{wp}$ (\%) &  & 10.2, 8.0 & 7.9, 8.3 & 4.7, 8.9 &  & 15.1, 8.4 & 12.6, 9.3 & 6.4, 8.9\\
\end{tabular}
\end{ruledtabular}
\end{table*}

\subsubsection{Alloys with $x\leq$0.4}

The alloys with $x\leq$0.4 have a behavior similar to that of Mn$_{3}$Sn$_{2}$.
Upon cooling from $T_{C1}$ to $T_{C2}$ there is a slight increase
in the intensity of a few peaks, namely (101), (011) and (210)+(112).
The appearance of the anti-$n$ (010) peak below $T_{C2}$ is concomitant
with a strong increase in the intensity of the (210)+(112) peaks.
Finally, the third transition at $T_{t}$ yields an increase in
the intensity of the (011) and (220)+(122)+(311) reflections together
with a reduction in the intensity of the (101) peak (Figure \ref{fig:I=00003Df(T)}).
The neutron diffraction patterns of this group of alloys were consequently
refined using the same sequence of magnetic structures than that prevailing
in Mn$_{3}$Sn$_{2}$. This resulted in satisfactory fits with low
$R$ magnetic ($R_{m}$) factors (4-12\%). Other attempts yielded
lower quality refinements with significantly higher $R$ factors.
Between $T_{C1}$ and $T_{C2}$, the alloys are simple ferromagnets
with the magnetic moments of both 4$c$ and 8$d$ sites pointing along
the $b$ axis ($F_{y}$ and $F_{By}$ modes, respectively, of
the magnetic space group $Pn'ma'$). We term this magnetic structure
Ferro-$b$. At $T_{C2}$, an antiferromagnetic component appears on
the 8$d$ site along the $a$ axis while the ferromagnetic component
of both sites is preserved. Hence, between $T_{C2}$ and $T_{t}$
the magnetic structure of the alloys with $x\leq$0.4 is still described
in the magnetic space group $Pn'ma'$ with a ferromagnetic 4$c$ sublattice
($F_{y}$ mode) and a canted ferromagnetic 8$d$ sublattice ($A_{Bx}F_{By}$
configuration). This configuration is named Canted Ferro-$b$. The
canting angle $\alpha$ weakly decreases upon increasing the Fe content
(from $\sim$50\textdegree{} to $\sim$47\textdegree)
but remains close to that in Mn$_{3}$Sn$_{2}$ ($\alpha$$\sim$51\textdegree).
Below $T_{t}$, the ferromagnetic component of both sites reorients
from the $b$ axis towards the $c$ axis. The low-temperature magnetic
structure, termed Canted Ferro-$c$, is then built upon a ferromagnetic
4$c$ sublattice ($F_{z}$ mode) and a canted ferromagnetic 8$d$
sublattice ($A_{Bx}F_{Bz}$), which implies a mixing of the magnetic
space groups $Pn'ma'$ and $Pn'm'a$. The canting angle is unchanged
at and below $T_{t}$. The refined magnetic moment components and
the main $R$ factors are given in Table \ref{tab:Mom 01 03}.

\subsubsection{Alloys with $x$=0.5 and 0.6}

\begin{figure}[h]
\begin{center}
\scalebox{0.60}{\includegraphics{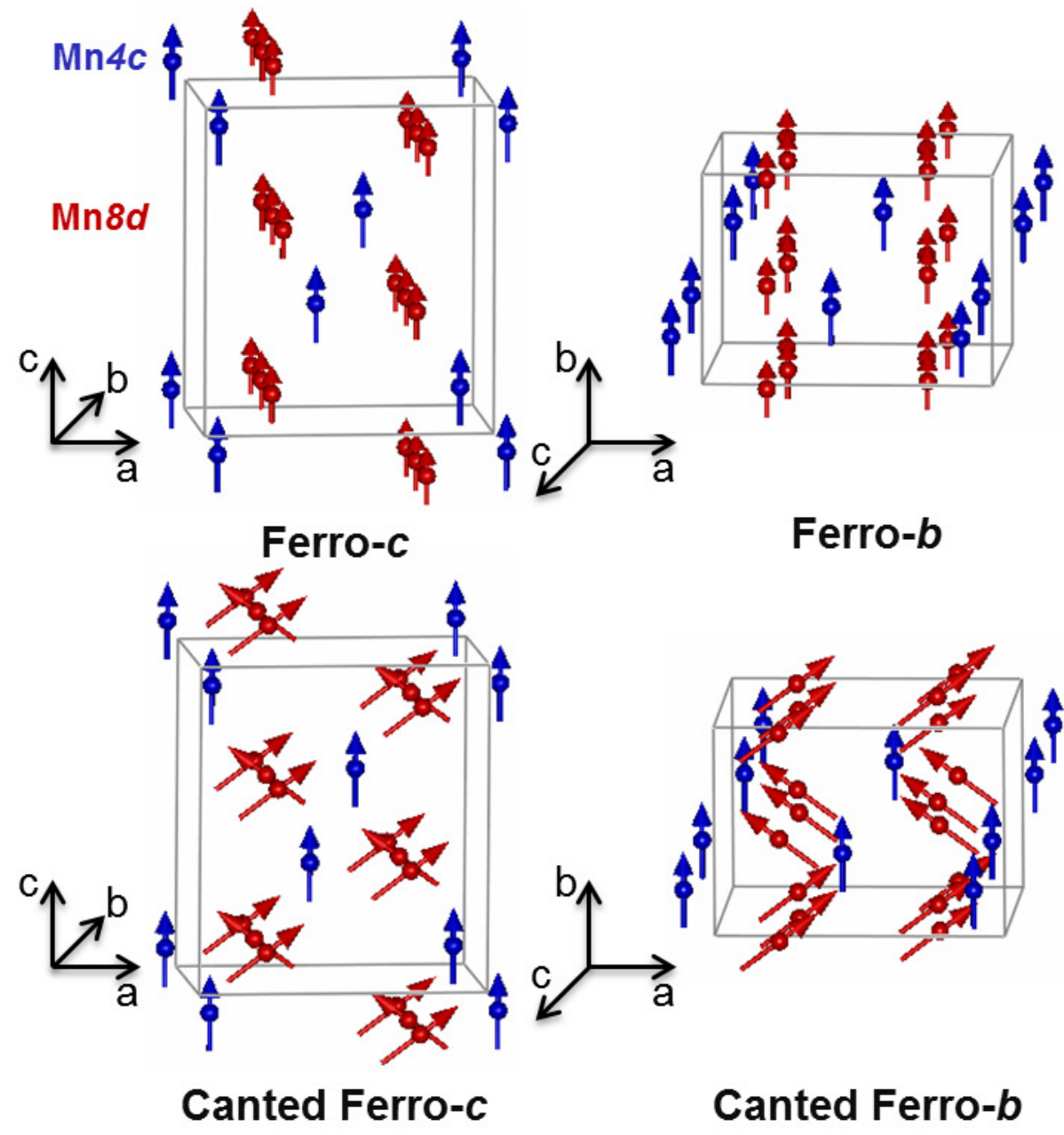}}
\end{center}
\vspace{-0.5cm}
\caption{\label{fig:StructMag}The four kinds of magnetic structure encountered
in the Mn$_{3-x}$Fe$_{x}$Sn$_{2}$ series.}
\vspace{-0.5cm}
\end{figure}

\begin{table*}
\caption{\label{tab:Mom 05 06}Refined magnetic moments for Mn$_{2.5}$Fe$_{0.5}$Sn$_{2}$
and Mn$_{2.4}$Fe$_{0.6}$Sn$_{2}$ from neutron data.}
\begin{ruledtabular}
\begin{tabular}{lcccccccc}
 &  & & $x$=0.5 &  &  &  & $x$=0.6 & \\
 &  & 230 K & 190 K & 2 K &  & 230 K & 130 K & 2 K\\
\colrule
Site 8$d$ & Mode & $F_{Bz}$ & $A_{Bx}F_{Bz}$ & $A_{Bx}F_{Bz}$ &  & $F_{Bz}$ & $A_{Bx}F_{Bz}$ & $A_{Bx}F_{Bz}$\\
 & $m_{x}(\mu_{B})$ &  & 0.64(2) & 1.76(2) &  &  & 1.52(2) & 1.67(2)\\
 & $m_{y}(\mu_{B})$ &  &  &  &  &  &  & \\
 & $m_{z}(\mu_{B})$ & 1.00(4) & 1.40(3) & 1.75(2) &  & 0.84(6) & 1.64(6) & 1.71(4)\\
\vspace{3mm}
 & $m(\mu_{B}$) & 1.00(4) & 1.54(3) & 2.48(3) &  & 0.84(6) & 12.24(5) & 2.48(3)\\
Site 4$c$ & Mode & $F_{z}$ & $F_{z}$ & $F_{z}$ &  & $F_{z}$ & $F_{z}$ & $F_{z}$\\
 & $m_{x}(\mu_{B})$ &  &  &  &  &  &  & \\
 & $m_{y}(\mu_{B})$ &  &  &  &  &  &  & \\
 & $m_{z}(\mu_{B})$ & 1.15(5) & 1.42(4) & 1.78(5) &  & 1.20(6) & 1.42(7) & 1.69(5)\\
\vspace{3mm}
 & $m(\mu_{B}$) & 1.15(5) & 1.42(4) & 1.78(5) &  & 1.20(6) & 1.42(7) & 1.69(5)\\
$R_{mag}$, $R_{wp}$ (\%) &  & 12.2, 8.2 & 9.4, 8.1 & 4.6, 10.1 &  & 12.9, 13.1 & 9.5, 10.1 & 5.6, 10.1\\
\end{tabular}
\end{ruledtabular}
\end{table*}

The alloys with $x$=0.5 and 0.6 behave differently. Beside the appearance
of the anti-$n$ (010) reflection below $T_{C2}$ and the concomitant
increase in the intensity of the (210)+(112) peaks, there is no further
perceptible anomaly and the intensity of the other magnetic lines
continuously increases upon cooling below $T_{C1}$ (Figure \ref{fig:I=00003Df(T)}).
This indicates that for $x$=0.5 and 0.6 there is no spin reorientation,
in agreement with the field-cooled magnetization data of reference~\onlinecite{Fe}. 
The best refinements (Table \ref{tab:Mom 05 06}) show
that upon cooling from $T_{C1}$ down to $T_{C2}$ both alloys
are simple collinear ferromagnets (Ferro-$c$ magnetic structure),
described in the magnetic space group $Pn'm'a$, with the magnetic
moments of both 4$c$ and 8$d$ sites pointing along $c$ axis ($F_{z}$
and $F_{Bz}$ modes, respectively). Below $T_{C2}$, the appearance
of the antiferromagnetic mode $A_{Bx}$ for the 8$d$ position yields
a canted ferromagnetic arrangement ($A_{Bx}F_{z}$) for this site
while the ferromagnetic configuration ($F_{z}$) of 4$c$ position
is kept. As is the case with the poorer Fe alloys, the magnetic space
groups $Pn'ma'$ and $Pn'm'a$ are mixed in the low temperature Canted
Ferro-$c$ magnetic structure of $x$=0.5 and 0.6. Compared with the
poorer Fe alloys, the canting angle is further reduced to reach $\alpha$$\sim$44\textdegree{}
in Mn$_{2.4}$Fe$_{0.6}$Sn$_{2}$. The refined magnetic moments are
given in Table \ref{tab:Mom 05 06}.

\subsubsection{Alloys with $x$$\geq$0.7}

In addition to the transitions at $T_{C1}$ and $T_{C2}$, the
thermal dependence of the intensities for the richest Fe alloys signals
the occurrence of a reorientation of the ferromagnetic components
at $T_{t}$. However, unlike the alloys with $x$$\leq$0.4, the transition
at $T_{t}$ yields a reduction in the intensity of the (011) and (220)+(122)+(311)
lines as well as an increase in the intensity of the (101) line (Figure
\ref{fig:I=00003Df(T)}). The best refinements (Table \ref{tab:Mom 07 08})
showed that these intensity changes correspond to a reorientation
of the ferromagnetic component of both sites from the $c$ axis towards
the $b$ axis upon cooling. Hence, between $T_{C1}$ and $T_{C2}$,
the alloys with $x$$\geq$0.7 adopt the Ferro-$c$ configuration with
magnetic moments of both sites oriented along the $c$ axis ($F_{z}$
and $F_{Bz}$ modes of $Pn'm'a$). The appearance of the antiferromagnetic
mode $A_{Bx}$ for the 8$d$ position at $T_{C2}$ leads to a canted
ferromagnetic arrangement ($A_{Bx}F_{Bz}$ configuration) while the
ferromagnetic alignment ($F_{z}$) of the 4$c$ sublattice is kept.
This Canted Ferro-$c$ structure, which implies a mixing of the magnetic
space groups $Pn'ma'$ and $Pn'm'a$, is stable over a small temperature
interval down to $T_{t}$, temperature at which the ferromagnetic
components reorient towards the $b$ axis leading to the Canted Ferro-$b$
structure which involves magnetic modes ($A_{Bx}F_{Bz}$ and $F_{z}$
for the 8$d$ and 4$c$ positions, respectively) which all pertain
to the magnetic space group $Pn'ma'$. The canting angle, which does
not change at $T_{t}$ and is insensitive to further temperature lowering,  weakly
decreases upon increasing the Fe content to reach $\alpha$$\sim$41\textdegree{} 
in Mn$_{2.2}$Fe$_{0.8}$Sn$_{2}$.

\begin{table*}
\caption{\label{tab:Mom 07 08}Refined magnetic moments for Mn$_{2.3}$Fe$_{0.7}$Sn$_{2}$
and Mn$_{2.2}$Fe$_{0.8}$Sn$_{2}$ from neutron data. }
\begin{ruledtabular}
\begin{tabular}{lccccccc}
 &  & \multicolumn{2}{c}{$x$=0.7}   &  &  & $x$=0.8 & \\
 &  & 230 K & 2 K &  & 230 K & 175 K & 2 K\\
\colrule
Site 8$d$ & Mode & $F_{Bz}$ & $A_{Bx}F_{By}$ &  & $F_{Bz}$ & $A_{Bx}F_{Bz}$ & $A_{Bx}F_{By}$\\
 & $m_{x}(\mu_{B})$ &  & 1.57(2) &  &  & 0.41(4) & 1.51(2)\\
 & $m_{y}(\mu_{B})$ &  & 1.75(7) &  &  &  & 1.75(4)\\ 
 & $m_{z}(\mu_{B})$ & 1.12(5) &  &  & 1.24(4) & 1.51(5) & \\
\vspace{3mm}
 & $m(\mu_{B}$) & 1.12(5) & 2.35(6) &  & 1.24(4) & 1.57(5) & 2.31(4)\\
Site 4$c$ & Mode & $F_{z}$ & $F_{y}$ &  & $F_{z}$ & $F_{z}$ & $F_{y}$\\
 & $m_{x}(\mu_{B})$ &  &  &  &  &  & \\
 & $m_{y}(\mu_{B})$ &  & 1.25(6) &  &  & 1.17(6) & 1.21(6)\\
 & $m_{z}(\mu_{B})$ & 1.15(6) &  &  & 0.81(7) &  & \\
\vspace{3mm}
 & $m(\mu_{B}$) & 1.15(6) & 1.25(6) &  & 0.81(7) & 1.17(6) & 1.21(6)\\
$R_{mag}$, $R_{wp}$ (\%) &  & 14.8, 11.3 & 12.3, 13.7 &  & 9.5, 9.1 & 8.1, 11.4 & 11.0, 7.7\\
\end{tabular}
\end{ruledtabular}
\end{table*}

\subsection{($x$, T) magnetic phase diagram}

\begin{figure}[h]
\begin{center}
\scalebox{0.68}{\includegraphics{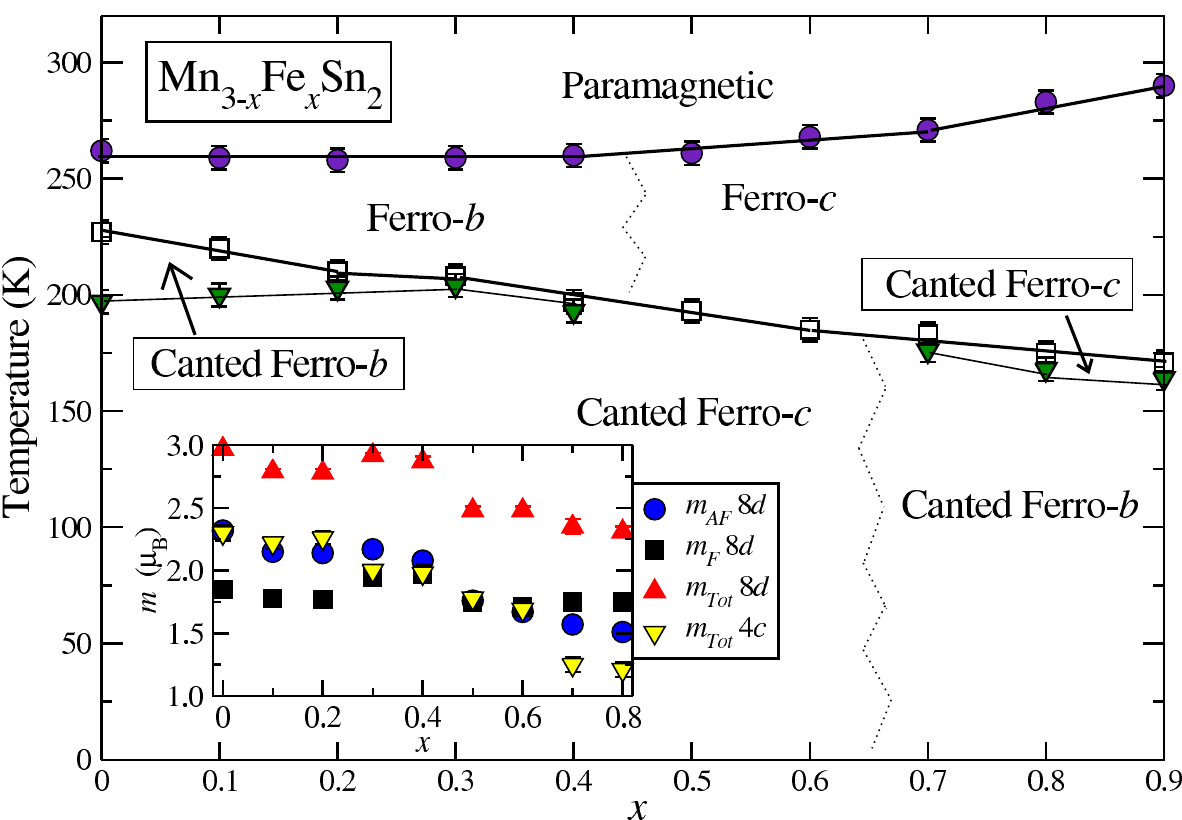}}
\end{center}
\vspace{-0.6cm}
\caption{\label{fig:diag}($x$, $T$) magnetic phase diagram of the Mn$_{3-x}$Fe$_{x}$Sn$_{2}$
series ($x$$\leq$0.9). The inset shows the composition dependence
of the transition metal magnetic moment for both sites. }
\vspace{-0.0cm}
\end{figure}

Combining the present data with those from references~\onlinecite{Mn3Sn2,Fe,PRB}
allows the construction of the ($x$, $T$) magnetic phase diagram as shown in Figure
\ref{fig:diag}. The thick lines correspond to second-order transitions
while thin lines are used for first-order transitions. The broken
lines mark uncertain boundaries. Depending on iron content and temperature,
four kinds of magnetic structure are identified: two ferromagnetic
arrangements with moments along either the $b$ axis or $c$ axis
(Ferro-$b$ and Ferro-$c$, respectively) and two canted ferromagnetic
configurations with ferromagnetic components along the $b$ axis or
$c$ axis (Canted Ferro-$b$ and Canted Ferro-$c$,
respectively).

As shown in the inset of Figure \ref{fig:diag}, the iron doping also
influences the magnitude of magnetic moments. The purely ferromagnetic
moment on 4$c$ site strongly decreases upon Fe substitution from
$\sim$2.3 $\mu_{B}$ in Mn$_{3}$Sn$_{2}$ down to $\sim$1.2 $\mu_{B}$
in Mn$_{2.2}$Fe$_{0.8}$Sn$_{2}$. Since the 4$c$ site is predominantly
populated by Fe atoms, this indicates that Fe carries a lower magnetic
moment than Mn in these phases, as often observed in $T$-$X$ intermetallic
compounds ($T$=Mn or Fe ; $X$=metalloid).\cite{MnSn2,FeSn2} Regarding the magnetic
moment on the 8$d$ site, our neutron refinements show that its ferromagnetic
component is almost constant throughout the series close to $\sim$1.8
$\mu_{B}$. At the same time, the antiferromagnetic component strongly
reduces upon doping from $\sim$2.3 $\mu_{B}$ in Mn$_{3}$Sn$_{2}$
down to $\sim$1.5 $\mu_{B}$ in Mn$_{2.2}$Fe$_{0.8}$Sn$_{2}$,
which results in a continuous reduction of the canting angle from
$\sim$51\textdegree{} to $\sim$40\textdegree{} (see Figure \ref{fig:dft_Total_magn}(a)).

\section{Electronic structure calculations\label{sec:Electronic}}

\begin{figure}[h]
\begin{center}
\scalebox{0.99}{\includegraphics{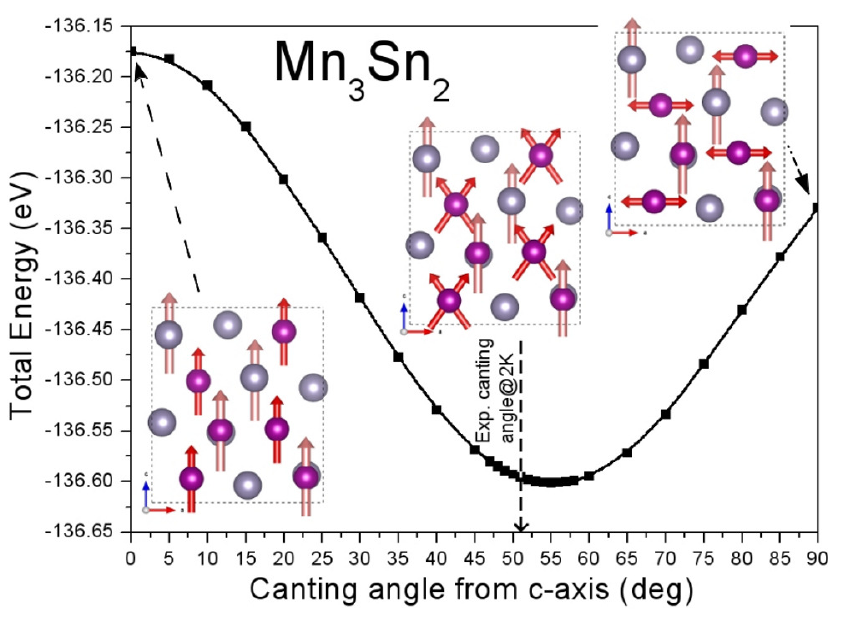}}
\end{center}
\vspace{-0.5cm}
\caption{\label{fig:dft-Canting}Change in total energy as a function of canting
angle in Mn$_{3}$Sn$_{2}$ resulted from DFT calculations. }
\vspace{-0.0cm}
\end{figure}

\begin{figure}
\begin{center}
\scalebox{0.99}{\includegraphics{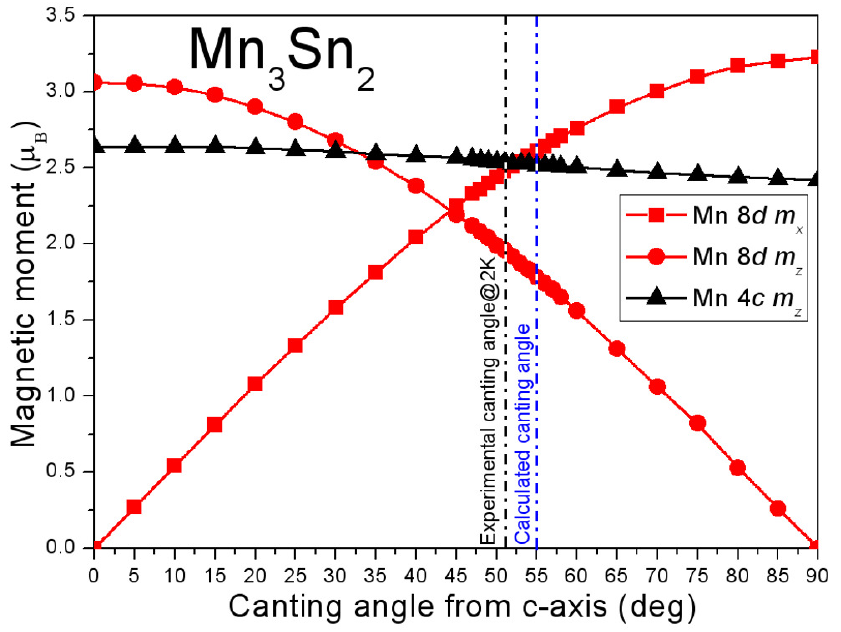}}
\end{center}
\vspace{-0.5cm}
\caption{\label{fig:dft-Mag_Mn3Sn2}Evolution of the site projected magnetic
moments with the canting angle in Mn$_{3}$Sn$_{2}$ resulted from
DFT calculations. }
\vspace{-0.5cm}
\end{figure}

 We first investigated the total energy dependence on the canting
angle $\alpha$ in the ground state (Canted Ferro-$c$ type magnetic
structure) of the Mn$_{3}$Sn$_{2}$ parent alloy as shown in Fig.
\ref{fig:dft-Canting}. The canting angle was constrained between
0 and 90\textdegree{} as displayed in the insets of Fig. \ref{fig:dft-Canting}.
The minimum in the total energy was found at $\alpha$=55\textdegree,
which is in a remarkable agreement with the value found experimentally
($\alpha$=51\textdegree). With the alteration of the canting angle,
the amplitude of the magnetic moments on the 8$d$ site projected
along the crystallographic axis also adjusts accordingly; $m_{z}$
continuously increases at the expense of $m_{x}$ as shown in Fig.
\ref{fig:dft-Mag_Mn3Sn2}. Around $\alpha$=55\textdegree, values
of $m_{x}$=2.61 $\mu_{B}$ and $m_{z}$=1.78 $\mu_{B}$ for Mn
$8d$ and $m_{z}$=2.50 $\mu_{B}$ for Mn $4c$ are calculated in
agreement with the experimental values reported in reference~\onlinecite{PRB}
($m_{x}$=2.32 $\mu_{B}$ and $m_{z}$=1.85 $\mu_{B}$ for Mn
$8d$ and $m_{z}$=2.30 $\mu_{B}$ for Mn $4c$).

Experimentally, a strong site preference of the iron dopant to occupy
the 4$c$ crystallographic site in the orthorhombic lattice has been
evidenced (see Sec. \ref{sub:Para} and Figure \ref{fig:Occ_Fe}).
This finding allows us to construct a simplified model based on a
supercell where we only consider the replacement of the Mn atoms by
Fe on the 4$c$ site. The magnetic moments reorient both with increasing
temperature ($x$$<$0.6) and with increasing Fe content ($x$$\geq$0.6)
from the Canted Ferro-$c$ structure into the Canted Ferro-$b$
one (Fig. \ref{fig:StructMag}). Therefore we have performed constrained
(angle) calculations for both spin arrangements. However, we have
found that the difference in total energies are vanishingly small
(in the range of $10^{-6}$ eV) regardless of the applied quantization
axis and it was not possible to converge the self-consistence solutions
within a precision high enough that confidently allow us to distinguish
the most stable orientation. Additionally, we find an
almost zero orbital contribution to the magnetic moment, thus a simple
picture of Mn being in a $d^{5}$ (Mn$^{2+}$) state with zero orbital
angular momentum can be drawn based on Hund\textquoteright s rule
with singly occupied $d$ levels. On the other hand, the variation
of total energy with canting represents an energy term that is two
orders of magnitude larger ($\sim$$10^{-4}$ eV) than the spin-orbit
coupling that suggests the exchange competition mainly contributing
to the peculiar magnetic order. 

\begin{figure}
\begin{center}
\scalebox{1.00}{\includegraphics{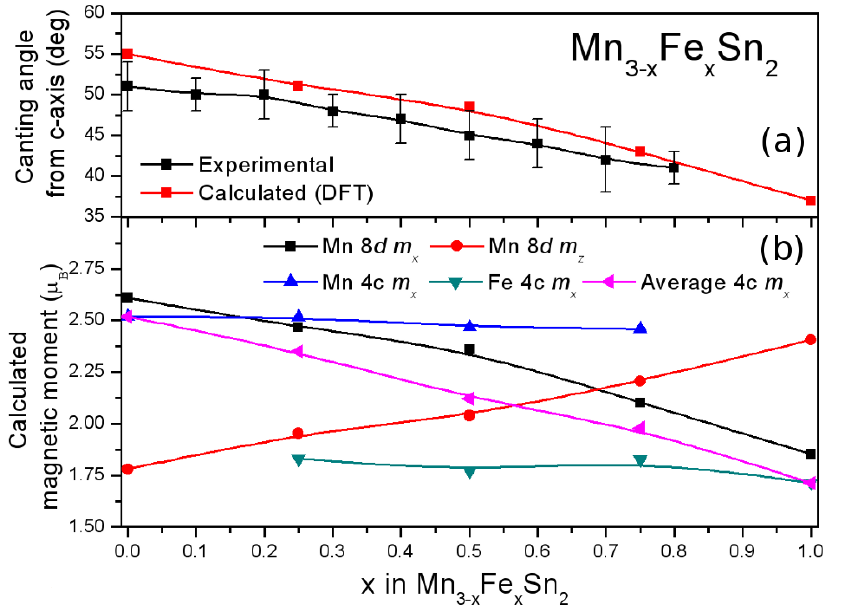}}
\end{center}
\vspace{-0.5cm}
\caption{\label{fig:dft_Total_magn}Evolution of the site projected magnetic
moments and canting angle in Mn$_{3-x}$Fe$_{x}$Sn$_{2}$ resulted
from DFT calculations.}
\vspace{-0.5cm}
\end{figure}

The comparison between the predicted and experimentally measured canting
angles with Fe doping is plotted in Fig. \ref{fig:dft_Total_magn}(a).
The calculated divergence of the magnetic moments from the quantization
axis with Fe content follows closely the experimentally established
values. As more Fe for Mn is introduced into the lattice, the canting
angle closes in with the magnetic moments leaning towards a more
ferromagnetic alignment. Fig.\ref{fig:dft_Total_magn}(b) shows the
site projected magnetic moments with Fe doping. The diminished magnetic
moments along the $a$ axis ($m_{x}$) on the 8$d$ site from 2.61
to 1.85 $\mu_{B}$ with Fe addition follows the simultaneous increase
of $m_{z}$ from 1.78 to 2.41 $\mu_{B}$. These changes reflect
well the trends observed experimentally (see inset of Figure \ref{fig:diag}),
though the increase in $m_{z}$ is not observed
experimentally and the computed magnetic moments for the doped alloys
tend to be higher than the experimental ones. These discrepancies
might be due, at least partially, to our supercell model which does
not take into account the weak occupancy of the 8$d$ site by Fe atoms.
Furthermore, the magnetic moment on the 4$c$ site is influenced slightly
by the doping and stays constant around 2.5 $\mu_{B}$ for Mn and
1.8 $\mu_{B}$ for the Fe atoms. Latter values do not seem to be
influenced by the canting,  suggesting strong ferromagnetic coupling
among them and little magnetic interaction between the two sublattices
which have been shown to magnetically order separately in Mn$_{3}$Sn$_{2}$.\cite{PRB}

\begin{figure}
\begin{center}
\scalebox{0.98}{\includegraphics{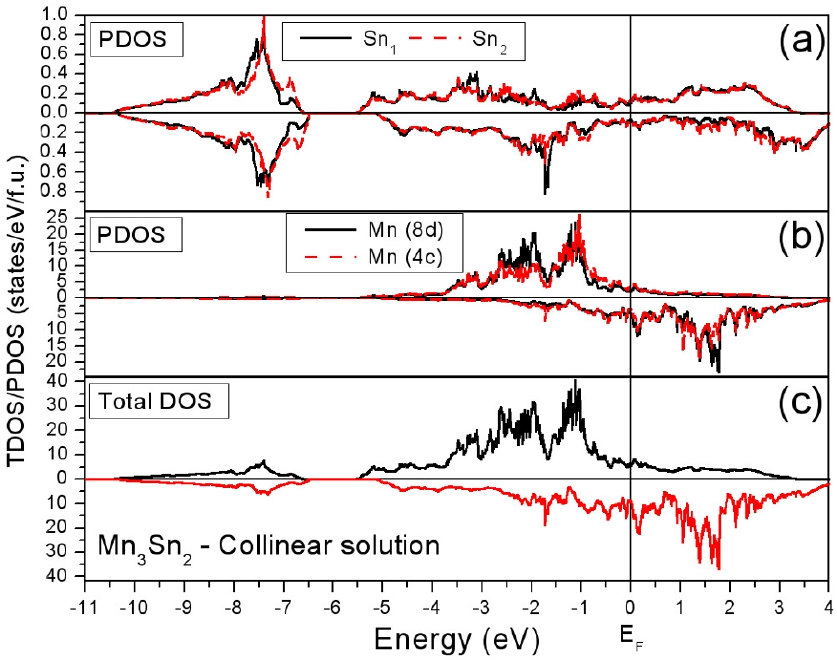}}
\end{center}
\vspace{-0.5cm}
\caption{\label{fig:PDOS}The partial density of states (PDOS) of Sn (a) and
Mn (b) in the ferromagnetic Mn$_{3}$Sn$_{2}$. The total spin-polarized
DOS is shown in figure (c).}
\vspace{-0.5cm}
\end{figure}

The partial density of states (PDOS) of Sn and Mn together with the
total DOS (TDOS) for the collinear (FM) configuration are shown in
Fig. \ref{fig:PDOS}. The metallic nature of the compound is apparent
as both Mn and Sn contribute with finite states to the electronic
structure, thus having bands crossing the Fermi level E$_{F}$. The
two inequivalent Sn sites have states distributed similarly in energy.
The $s$-states of Sn atoms show peaks at lower energy ranges of around
-7 eV, whilst electrons with both $p$- and $d$-character are continuously
present from about -5 eV to well above E$_{F}$. The presence of the
peak in the majority electron spin density around -3.2 eV that is
only mirrored in the minority spin channel at around -1.7 eV points
to a magnetic character of the Sn atoms. Indeed, the site-projected
magnetic analysis reveals -0.06 $\mu_{B}$ and -0.04 $\mu_{B}$ moments
for the two Sn sites caused by the proximity of large magnetic moments
on the transition metal sites. These hybridization induced Sn magnetic
moments are responsible for the large hyperfine fields measured by
$^{119}$Sn M\"ossbauer spectroscopy in Mn$_{3}$Sn$_{2}$.\cite{PRB}
PDOS of the Mn-sites (Fig. \ref{fig:PDOS}b) indicates strongly exchange
split bands. The energy landscape of the states is very similar for
both Mn sites. We can identify a broad bunch of strongly hybridized
peaks in the majority electron spin states between -3 and -1.6 eV.
In this energy range, Mn atoms at the 8$d$ site have significantly
higher contribution than that from the 4$c$. On the other hand, similar
number of states is observed between -1.6 and -0.8 eV from both crystallographic
sites. Latter bands are significantly narrower suggesting less
hybridization compared to the ones mentioned above. The exchange split
peaks in the minority electron spin states appear well above the Fermi
level. Considering the 8$d$ site dominated double peaks in the unoccupied
states at around 1.7 eV, we estimate a very strong exchange energy
between about 3.3 to 4 eV. On the other hand, the exchange energy
related to the 4$c$ site is considerably smaller around 2.2 to 3 eV.

\begin{figure}
\begin{center}
\scalebox{0.99}{\includegraphics{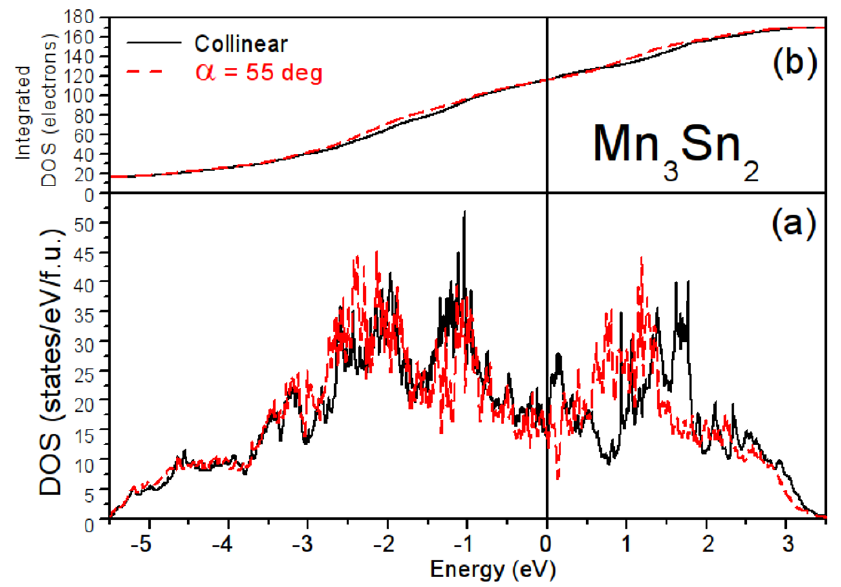}}
\end{center}
\vspace{-0.5cm}
\caption{\label{fig:TDOS} Non-collinear theory allows the comparison of the
total density of states (TDOS) using {[}N$_{tot}$=N$_{\uparrow}$+N$_{\downarrow}${]}.
Bottom panel (a) compares the NC (dashed line) and collinear FM TDOS (full line) of Mn$_{3}$Sn$_{2}$
and top panel (b) is the integrated TDOS of the same. }
\vspace{-0.5cm}
\end{figure}

We now turn our attention to a qualitative comparison of the significant
differences of the DOS between the FM and non-collinear ($\alpha$=55\textdegree) 
spin arrangements. Using non-collinear theory, we no
longer distinguish spin-up and spin-down channels, therefore the total
density of states in Fig. \ref{fig:TDOS} is calculated as {[}N$_{tot}$=N$_{\uparrow}$+N$_{\downarrow}${]}
for the FM state in order to compare the two magnetic solutions. In
other words, the collinear solution (black line) is another representation
of the results shown in Fig. \ref{fig:PDOS}(c). The NC spin arrangement
(red) causes the 8$d$ site dominated double peaks in the unoccupied
states (1.7 eV) to lower by about 0.4 eV, suggesting that smaller
exchange energy is required to maintain the NC configuration. On the
other hand, the occupied number of states between -1.6 and -0.8 eV,
where both Mn sites have dominant contribution is reduced. However
these states re-appear on the -3 and -1.6 eV energy range, where the
electronic states at the 8$d$ site is dominant. These changes are
also reflected in the integrated DOS in Fig. \ref{fig:TDOS} (top).
From E=-3 eV, more occupied states are present for the NC spin arrangement
than for the FM that lowers the total energy of the NC solution. In
addition, the calculated total density of states at E$_{F}$ {[}N$_{tot}$(E$_{F}$)=N$_{\uparrow}$+N$_{\downarrow}${]}
for the collinear FM case is (14.3 states/eV/f.u.) slightly higher
than that for the non-collinear case (13.4 states/eV/f.u.). A stability
criteria, whereby the high number of states can cause ferromagnetism
unstable and promote a non-collinear AFM arrangement was reported
earlier in similar Mn-based orthorhombic systems. \cite{Sredniawa2001266,Zach,Alex_PRL,Zsolt_PRB}
The tipping point, where FM density of states at E$_{F}$ becomes
lower than that of the NC one is just bellow the E$_{F}$ by about
-0.003 eV ($\lesssim$$k_{B}T$), that could explain the occurrence
of NC to FM transition at finite temperatures. 

\begin{figure}
\begin{center}
\scalebox{0.99}{\includegraphics{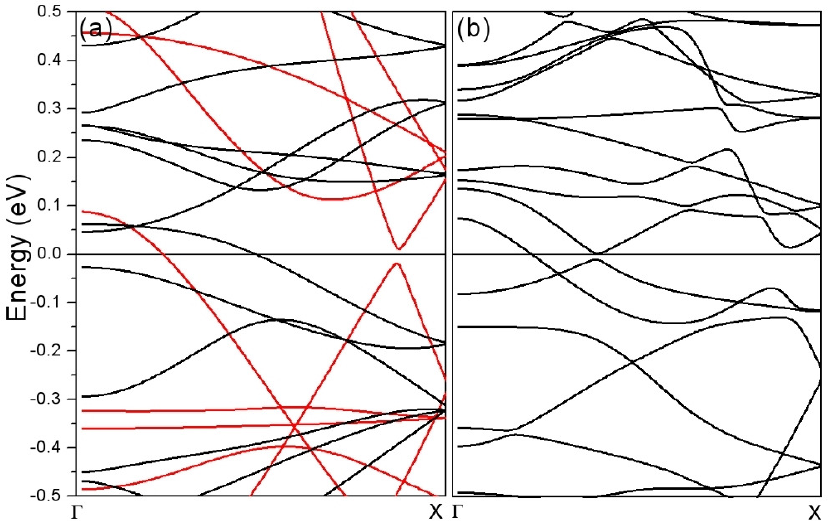}}
\end{center}
\vspace{-0.5cm}
\caption{\label{fig:Band} Band dispersion of the FM solution (a) and the NC
spin arrangement (b) calculated for Mn$_{3}$Sn$_{2}$. A hybridization
gap appears for the NC spin arrangement reducing the number of bands
that cross the Fermi level lowering the total energy through the band
mechanism \cite{PRL_Band}.}
\vspace{-0.5cm}
\end{figure}

The comparison of the band dispersion relations in Fig. \ref{fig:Band}
indicates further distinctiveness between the two magnetic solutions
related to the electronic structures around E$_{F}$. In the NC case
(right panel), some bands are pushed lower in energy as a result of
less exchange splitting compared to the FM solution (left panel),
in line with the observations from the DOS plots (Fig. \ref{fig:TDOS}).
This feature is most apparent close to the $X$ point around the Fermi
level. Another important consequence of the non-collinear spin arrangement
is the opening of a hybridization gap around half way from the $\Gamma$
to $X$ point. This mechanism, where the bands hybridize with one
another has been identified to stabilize non-collinearity. \cite{PRL_Band,IrMnSi}
Based on the analysis of Liz\`arraga $et$ $al.$,\cite{PRL_Band} the total energy can be
lowered if a gap opens at the Fermi level by the hybridization of
the orthogonal spin-up and spin-down states. Many new states contributing
to this energy lowering mechanism can be obtained by a non-collinear
spin arrangement, if there is a nesting between spin-up and spin-down
Fermi surfaces at the ferromagnetic configuration. If the hybridization
causes an opening of the band gaps around E$_{F}$,  the number
of bands that cross it is reduced and the total energy of the system
is lowered. Based on these observations, we can conclude that Mn$_{3}$Sn$_{2}$
also belongs to the class of non-collinear materials stabilized by this band mechanism.

\section{Synchrotron diffraction experiments\label{sec:Synchrotron}}

In the framework of Representation Analysis,\cite{Bertaut}
for describing the Canted Ferro-$c$ magnetic structure, which involves
the $F_{z}$, $F_{Bz}$ and $A_{Bx}$ modes, it is necessary to mix
the two orthorhombic magnetic space groups $Pn'ma'$ and $Pn'm'a$.
As discussed in reference~\onlinecite{PRB} for Mn$_{3}$Sn$_{2}$, within
which this canted state is stabilized below $T_{t}$, two possible
mechanisms have been identified to explain this uncommon situation.
The first one requires the presence of unusually large high-order
non-Heisenberg terms\cite{non-Heisenberg} in the spin Hamiltonian.\cite{Bertaut}
The other one implies the occurrence of a weak (undetected from neutron
diffraction) monoclinic distortion. The magnetic space
groups $Pn'ma'$ and $Pn'm'a$ have the monoclinic magnetic space
group $P2'_{1}/n'11$ (notice the non-conventional setting) as intersection.
In the monoclinic space group $P2_{1}/n11$, the Mn atoms occupy three
4$e$ positions. The basis functions $A_{x}$, $F_{y}$ and $F_{z}$
all belong to the relevant irreducible representation of $P2_{1}/n11$ (Table \ref{tab:The-magnetic-modes}).
Therefore, if the distortion is present, it is then not necessary
to mix different monoclinic magnetic space group to describe the Canted
Ferro-$c$ magnetic structure.

\begin{figure}
\begin{center}
\scalebox{0.48}{\includegraphics{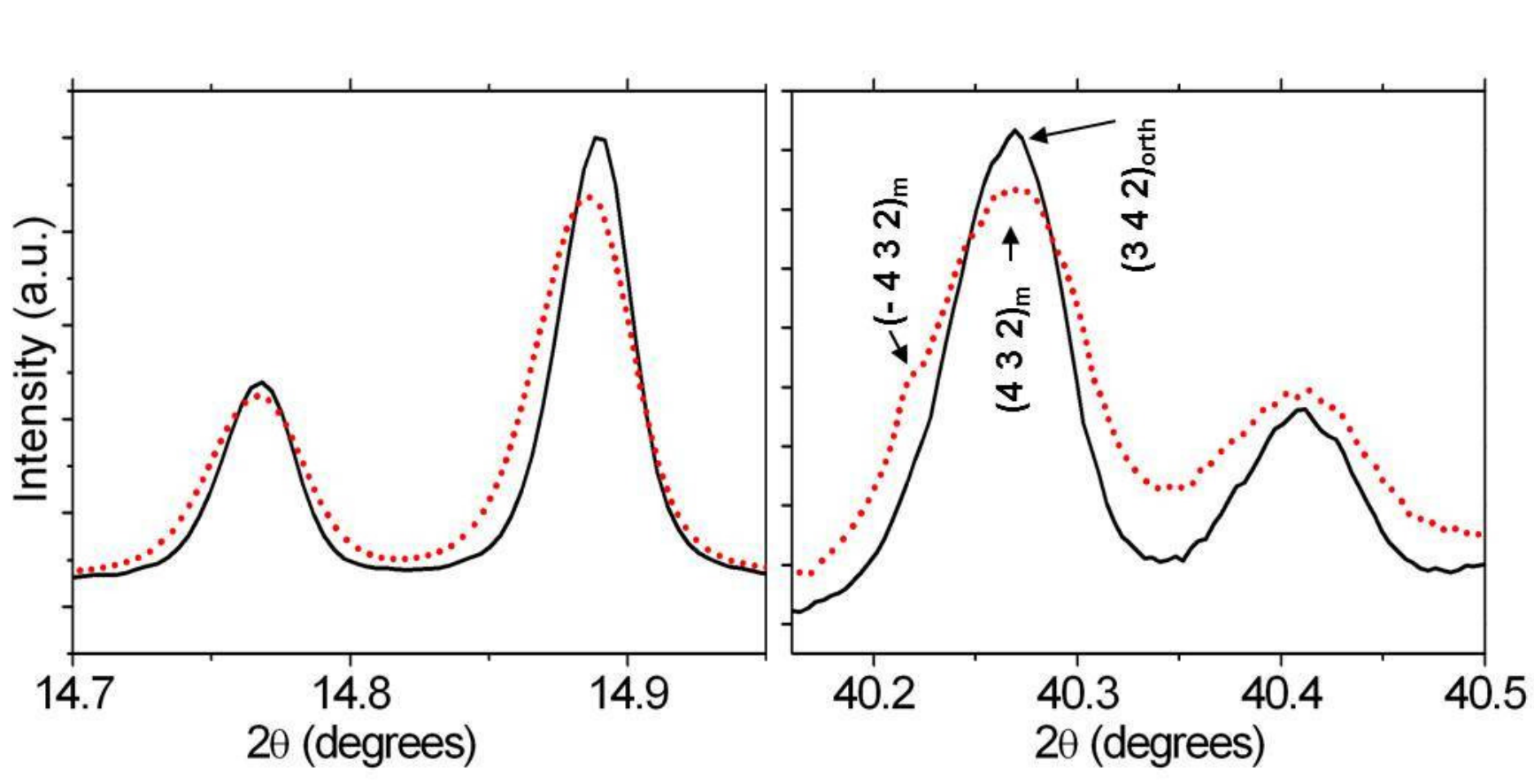}}
\end{center}
\vspace{-0.7cm}
\caption{\label{fig:Enlargment}Some selected angular regions of the synchrotron
diffraction patterns of Mn$_{3}$Sn$_{2}$ at 300 K (full line) and 175 K (dotted line) showing
the broadening of the diffraction peaks upon cooling. The 175 K pattern
has been shifted to allow for comparisons.}
\vspace{-0.5cm}
\end{figure}

To check the latter possibility, we have undertaken
X-ray diffraction experiments using synchrotron radiation on the Mn$_{3}$Sn$_{2}$
parent compound in the 175-300 K temperature range. The Mn$_{3}$Sn$_{2}$ sample presents a good crystallinity
as the half-width measured at 300 K is 0.038\textdegree{} at 2$\theta$,
close to the instrumental resolution (see Section \ref{sec:Introduction}).  At 300 K, the
line broadening was interpreted from microstructural effects (size
and microstrain). A weak supplementary broadening of the
diffraction peaks is observed upon cooling (Figure \ref{fig:Enlargment}).
This effect, which cannot be ascribed to crystallites size change,
might be indicative of a lowering of symmetry although we did not
find any clear evidence of marked splitting down to 175 K (the lowest
temperature of the experiment) but instead a shouldering of some diffraction lines 
at high enough diffraction angles (Figure \ref{fig:Enlargment}). 

\begin{figure}
\begin{center}
\scalebox{0.32}{\includegraphics{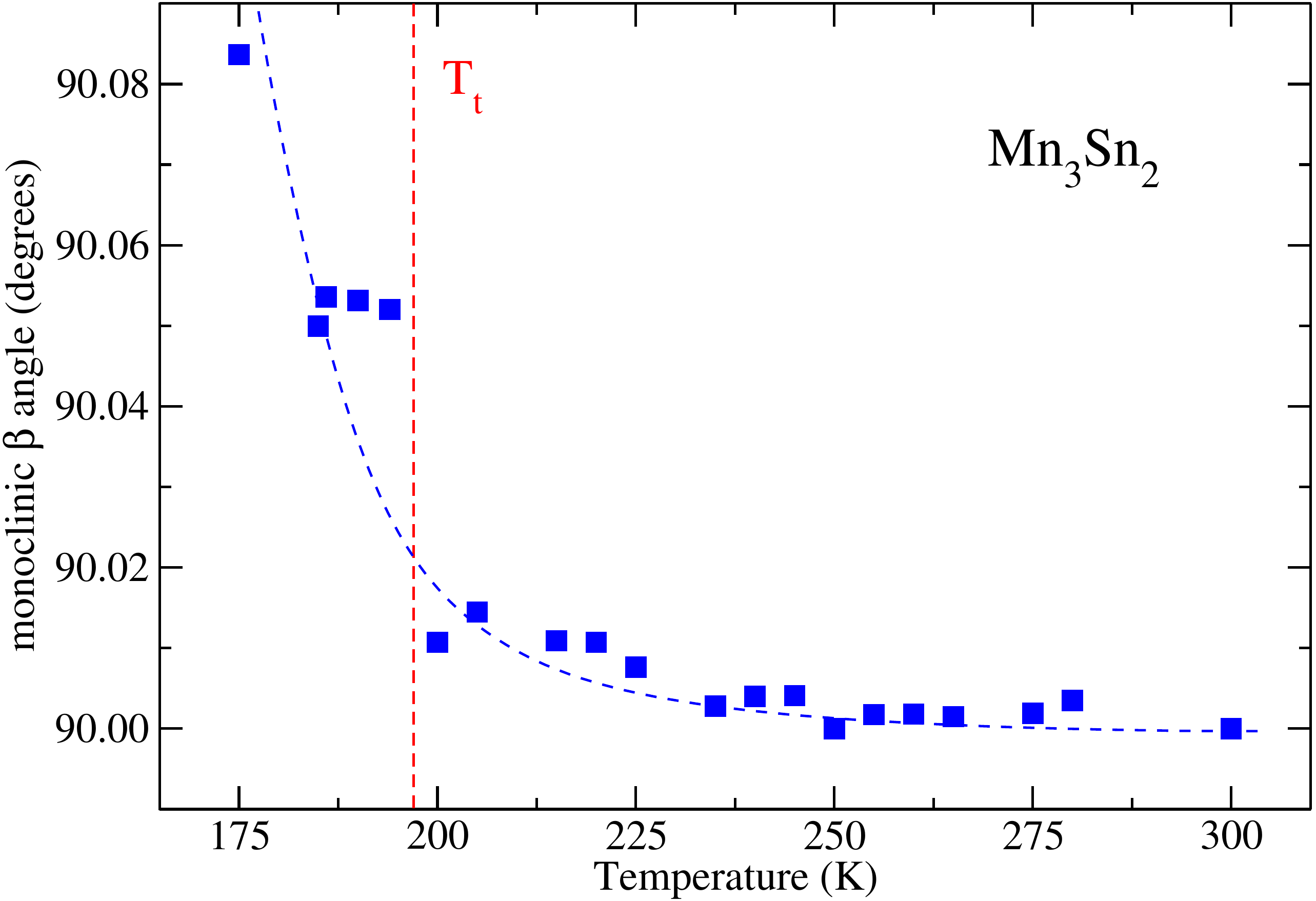}}
\end{center}
\vspace{-0.5cm}
\caption{\label{fig:beta}Temperature dependence of the monoclinic angle in
Mn$_{3}$Sn$_{2}$ from synchrotron data. The dashed line is a guide
to the eye. The vertical line marks the $T_{t}$ transition temperature
as determined from magnetic data. \cite{Mn3Sn2}}
\vspace{-0.5cm}
\end{figure}

We first tested refining a few  synchrotron diffraction patterns 
using the three possible monoclinic subgroups of $Pnma$:
$P12_{1}/m1$, $P12_{1}/n1$, and $P12_{1}/a1$.
A total of 22 parameters was refined for the 300 K pattern: 1 for the
zero-shift, 9 for the crystallites size, 9 for the strains, 3 for the
lattice (the monoclinic $\beta$ angle was kept fixed at 90\textdegree).
The refined 300 K pattern was used as a reference for those recorded
at lower temperature. The 18 (9+9) microstructural parameters were
kept fixed at their refined 300 K values while the $\beta$ angle
was let free to vary. Consequently, only 5 parameters were refined
for temperature below 300 K: the zero-shift and the 4 lattice parameters
($a_{mono}$, $b_{mono}$, $c_{mono}$, and $\beta$). The $P12_{1}/m1$ space group 
does not allow to take into account the peak broadening upon cooling contrary to the 
$P12_{1}/n1$ and $P12_{1}/a1$ space groups. 
$P12_{1}/n1$  leads however to significantly better fits than $P12_{1}/a1$  ($\chi^2$$\sim$2.0 $vs.$ $\sim$2.6).
The synchrotron diffraction patterns were therefore all refined using the  $P12_{1}/n1$ space group 
($a_{mono}\equiv b_{ortho}$ ; $b_{mono}\equiv a_{ortho}$ ; $c_{mono}\equiv c_{ortho}$). In all cases,
$R_{p}$ and $R_{wp}$ factors converge satisfactorily to values lower
than 5\%. 
The temperature dependence of the refined monoclinic
$\beta$ angle is presented in Figure \ref{fig:beta}. Upon cooling,
the $\beta$ angle remains equal to 90\textdegree{} down to  about 200
K, temperature below which it increases to reach $\beta$=90.083(1)\textdegree{} 
at 175 K. Thus there is a weak but clear monoclinic distortion of
the orthorhombic cell of Mn$_{3}$Sn$_{2}$ for temperatures below
the spin reorientation temperature  ($T_{t}$$\sim$197
K) previously evidenced by neutron
diffraction and $^{119}$Sn M\"ossbauer spectroscopy.\cite{PRB} Based on these new evidences, we can conclude that
the Canted Ferro-$c$ magnetic structure of Mn$_{3}$Sn$_{2}$ occurring
below $T_{t}$ can be explained by a weak monoclinic distortion and therefore 
is not due to the presence of unusually large high-order terms in
the spin Hamiltonian. Experiments performed down to lower temperatures,
and preferentially with an even higher resolution diffractometer,
should reveal the full magnitude of the distortion. It is likely that
this distortion holds in Mn$_{3-x}$Fe$_{x}$Sn$_{2}$ with $x$$\leq$0.6,
which all adopt the Canted Ferro-$c$ magnetic structure at low temperature
(see Figure \ref{fig:diag}). For $x$$\geq$0.7 this magnetic arrangement
is only stabilized over a restricted interval at intermediate temperatures.
It could be interesting to verify how the iron substitution alters
the cell distortion and, in particular, if for $x$$\geq$0.7 the monoclinic distortion
manifests only in an intermediate temperature range before to disappear
upon further cooling, as previously observed for instance in YVO$_{3}$.\cite{YVO3}

\section{Summary \label{sec:Summary}}

We have studied the magnetocaloric Mn$_{3-x}$Fe$_{x}$Sn$_{2}$ alloys
($x$$\leq$0.8) using neutron powder diffraction and DFT calculations.
In addition, the Mn$_{3}$Sn$_{2}$ parent compound has been examined
by synchrotron diffraction. We found that the Mn$_{3-x}$Fe$_{x}$Sn$_{2}$
alloys can be classified into three groups according to the thermal
evolution of their magnetic structures. The alloys with $x$$\leq$0.4
behave similarly to Mn$_{3}$Sn$_{2}$: they order in the Ferro-$b$
arrangement at $T_{C1}$, adopt the Canted Ferro-$b$ configuration
between $T_{C2}$ and $T_{t}$ before the ferromagnetic components
reorient at $T_{t}$ to yield the Canted Ferro-$c$ structure. The
richest Fe alloys ($x$=0.7 and 0.8) show an almost reverted sequence
as they order in the Ferro-$c$ structure at $T_{C1}$ and evolve
into the Canted Ferro-$c$ arrangement over a limited temperature
interval between $T_{C2}$ and $T_{t}$, a temperature below which they
adopt the Canted Ferro-$b$ configuration. There is no spin reorientation
for the intermediate iron contents ($x$=0.5 and 0.6) which order
in the Ferro-$c$ structure below $T_{C1}$ and transform directly
towards the Canted Ferro-$c$ arrangement at $T_{C2}$. The Fe atoms
are found to preferentially localize on the 4$c$ transition metal
site whose magnetic moment decreases upon increasing $x$, suggesting
lower magnetic moment on Fe atoms than on Mn atoms in these phases
as confirmed by DFT computations. The magnetic moments and canting
angle calculated using non-collinear computations agree well with
the experimental data throughout the series. The comparison with collinear
theory indicates that the canted ferromagnetic states are stabilized
by the opening of a hybridization gap. Temperature dependent synchrotron
diffraction experiments on Mn$_{3}$Sn$_{2}$ performed in the 175-300 K 
temperature range show that its cell undergoes a weak 
monoclinic distortion below $T_{t}$ ({$\beta$$\sim$90.08\textdegree{} 
at 175 K). Consequently, it is unnecessary to invoke
fourth or higher order terms in the spin Hamiltonian to explain the
Canted Ferro-$c$ magnetic structure since in monoclinic symmetry
the two involved $F_{z}$ and $A_{x}$ magnetic modes both pertain
to the same irreducible representation.

\section*{Acknowledgments}

We are indebted to the Institut Laue Langevin for the provision of
research facilities. We are grateful to our local contact (S. Capelli)
for her help during the recording of the data. Financial support is
acknowledged from EPSRC grant EP/G060940/1 (Z.G.). Computing resources
provided by Darwin HPC and Camgrid facilities at The University of
Cambridge and the HPC Service at Imperial College London are gratefully
acknowledged. Finally, inset of Fig. \ref{fig:dft-Canting} was prepared
using VESTA open-source software. \cite{Momma}

\end{document}